\documentclass[final,5p,times,twocolumn]{ms}

\usepackage{amssymb,amsfonts,amsmath,amstext,amsgen,amsopn,amsxtra,indentfirst,lscape,xtab}
\usepackage{multicol}
\usepackage{blindtext}
\usepackage{color}
\usepackage{epstopdf}

\newcommand{\arcmin}{$^{\prime}$}
\newcommand{\arcsec}{$^{\prime\prime}$}
\newcommand{\masyr}{mas\,yr$^{-1}$}
\newcommand{\kms}{km\,s$^{-1}$}
\newcommand{\dgr}{$^\circ$}

\newcommand{\vimos}{\texttt{VIMOS}}
\newcommand{\vimosifu}{\texttt{VIMOS-IFU}}

\journal{Science, February 7, 2013}

\begin{document}


\begin{frontmatter}

\title{\begin{flushleft}
{\sc \small Published in Science Express, February 14, 2013}\\
\end{flushleft}
\vspace{0.2cm}
An Integral View of Fast Shocks around Supernova 1006}

\author[mpia]{Sladjana Nikoli\'{c}}
\ead[mpia]{sladja.ast@gmail.com}
\author[mpia]{Glenn van de Ven}
\address[mpia]{Max Planck Institute for Astronomy, K\"{o}nigstuhl 17, D-69117, Heidelberg, Germany}

\author[bern]{Kevin Heng}
\address[bern]{University of Bern, Center for Space and Habitability, Sidlerstrasse 5, CH-3012, Bern, Switzerland}

\author[aip]{Daniel Kupko}
\author[aip]{Bernd Husemann}
\address[aip]{Leibniz Institute for Astrophysics Potsdam (AIP), An der Sternwarte 16, D-14482 Potsdam, Germany}

\author[cfa]{John C. Raymond}
\address[cfa]{Harvard-Smithsonian Center for Astrophysics, 60 Garden Street, Cambridge, MA 02138, U.S.A.}

\author[rutgers]{John P. Hughes}
\address[rutgers]{Department of Physics and Astronomy, Rutgers University, 136 Frelinghuysen Road, Piscataway, NJ 08854, U.S.A.}

\author[iac]{Jes\'{u}s Falc\'{o}n-Barroso}
\address[iac]{Instituto de Astrof\'{i}sica de Canarias, V\'{i}a L\'{a}ctea, E38205, La Laguna (Tenerife), Spain}

\end{frontmatter}


\textbf{Supernova remnants are among the most spectacular examples of astrophysical pistons in our cosmic neighborhood.  The gas expelled by the supernova explosion is launched with velocities $\sim 1000$ km s$^{-1}$ into the ambient, tenuous interstellar medium, producing shocks that excite hydrogen lines.
We have used an optical integral-field spectrograph to obtain high-resolution spatial-spectral maps that allow us to study in detail the shocks
in the northwestern rim of supernova 1006. The two-component H$\alpha$ line is detected at 133 sky locations. Variations in the broad line widths and the broad-to-narrow line intensity ratios across tens of atomic mean free paths suggest the presence of suprathermal protons, the potential seed particles for generating high-energy cosmic-rays.}

Supernova remnants, the expanding shells of material created in a stellar explosion, are astrophysical laboratories for studying non-thermal physics and high-velocity shocks, and have been scrutinized over a broad range of wavelengths.  The signatures of non-thermal electrons are typically manifested in the X-ray, $\gamma$-ray and radio range of wavelengths \cite{reynolds08}.  Complementary to these observations, hydrogen emission from ``Balmer-dominated shocks" \cite{cr78,bl79,ckr80}, around supernova remnants of thermonuclear origin, directly probe the proton populations \cite{heng10}.  Until now, all studies of Balmer-dominated shocks have typically used conventional spectrographs that yield rich spectral information but limited spatial information. Here we report on Balmer-dominated shocks around a supernova remnant using integral-field unit (IFU) spectroscopy, a technique that produces a three-dimensional ``data cube": two dimensions of space (across the sky) and a spectral dimension.  We selected supernova (SN) 1006 as our target, because it has a long history of serving as a laboratory for studying non-thermal physics and high-velocity shocks \cite{acero10,koyama95,bkv02,win03,rot04,cc08,petruk09}.\\

Long-slit spectroscopy, utilized in previous studies of Balmer-dominated shocks \cite{ghava02}, cumulatively measures the
H$\alpha$ emission emanating from regions much larger than
the characteristic length scale: the mean free path for interactions
between hydrogen atoms and electrons or ions, $L_{\rm mfp} \sim 1/n \sigma_{\rm ce}$ (with $n$ being the
pre-shock number density and $\sigma_{\rm ce} \sim 10^{-15}$ cm$^2$ denoting the typical cross section for
charge exchange).  Using the inferred range of densities for the northwestern rim of SN 1006 of
$n$ from 0.15 to $0.40$\,cm$^{-3}$ \cite{ray07,long03,acero07}, we estimated that the $0.^{\prime\prime}67$ size of a pixel in our
IFU observations corresponds to $\sim$\,5$L_{\rm mfp}$.
The high spatial resolution allowed us to separate out the contributions of bulk motion versus thermal velocity
to the measured line widths, an issue which has limited the interpretation of previous observations. It also enabled us to study the spatial variation of the line widths and ratios and hence changes in microphysics of a shock across
several atomic mean free paths. To buttress the second point, we have intentionally chosen a field of view that
zooms in on a region of simple geometry, implying that any substantial spatial variation detected cannot be due to variations
in density caused by geometric or projection effects.

\begin{figure*}[!t]
\begin{center}
\hspace{0.6in}
\includegraphics[width=0.7\textwidth,angle=-90,trim=0 20mm 0 0]{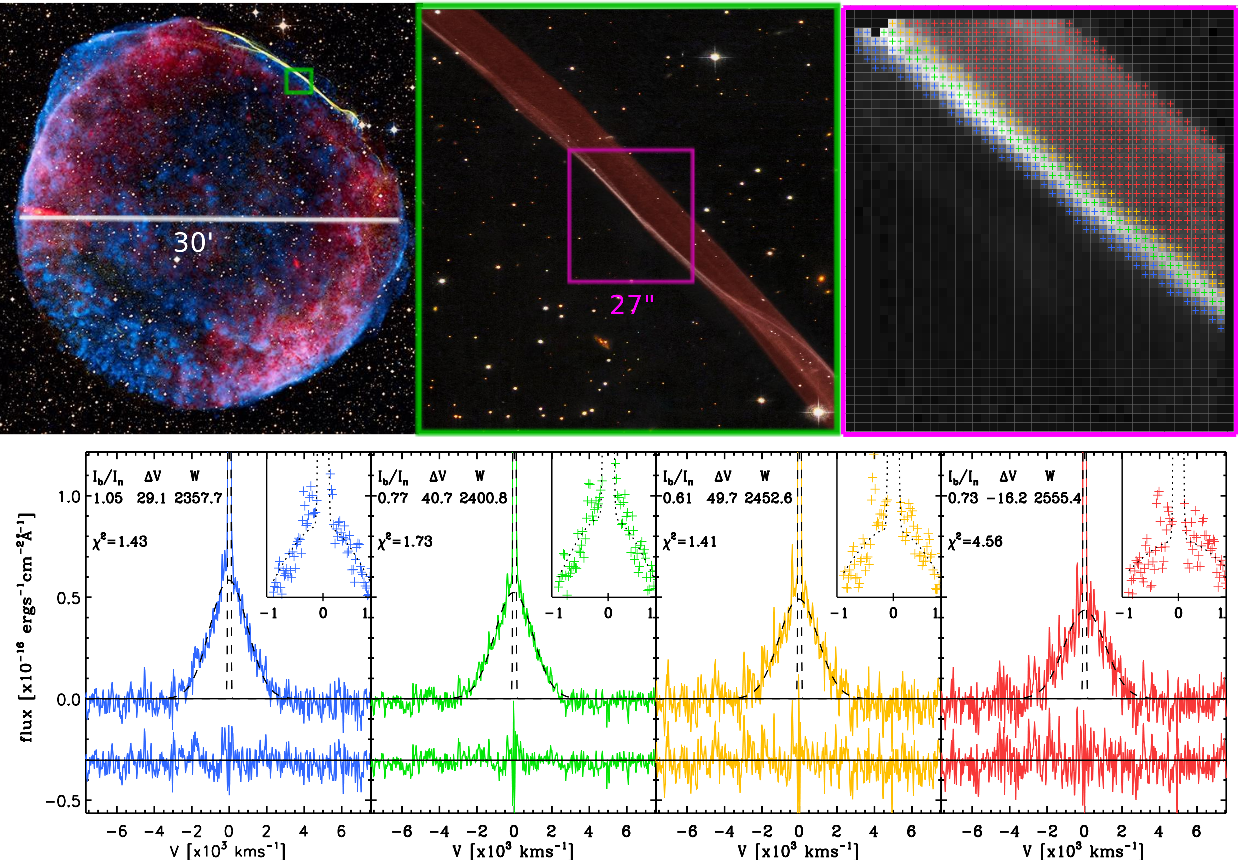}
\end{center}
\vspace{-0.2in}
\caption{\vimosifu\ spectroscopy of the shock front in the remnant of SN\,1006. The \emph{top-left} panel shows a
composite image of the full remnant ($\approx$30\arcmin\ in diameter), combining data from the Very Large Array
and Green Bank Telescope (red; NRAO/AUI/NSF/GBT/VLA/Dyer, Maddalena \& Cornwell), Chandra X-ray Observatory (blue;
NASA/CXC/Rutgers/G.\ Cassam-Chena\"i, J. Hughes et al.), 0.9\,m Curtis Schmidt optical telescope (yellow;
NOAO/AURA/NSF/CTIO/Middlebury College/F.\ Winkler), and Digitized Sky Survey (orange and light blue stars).
The green box indicates the region covered by the Hubble Space Telescope (HST) H$\alpha$ narrow-band image shown in the
\emph{top-middle}, with subsequently the magenta box indicating the region observed with the \vimosifu.
The \emph{top-right} panel shows the reduced data cube collapsed in wavelength around the H$\alpha$-line, recovering
the shock front. The crosses with four different colors indicate the pixels for which the spectra have been combined
to produce the spectra shown in the four panels at the \emph{bottom}.  In each of the bottom panels, the dashed black
lines show the best-fit double-Gaussian, with parameters given in the legenda: the intensity ratio of the broad to narrow
component $I_b/I_n$, the velocity offset between the broad and narrow line centroids $\Delta V$ (in \kms), and width of
the broad component $W$ (in \kms). The reduced $\chi^2$ values above unity along with the differences between the observed
spectra and their best-fits (shown below the spectra with an offset of -0.3 for clarity), indicate that non-Gaussianity is present.
Most of the reduced $\chi^2$ values above unity come from the
mismatching near the line core, as can be seen from the zoomed-in region (-1000,1000) \kms\ in the top-right corner of
bottom panels. On the horizontal axis is shown only the fitted region of the spectra,
while the y axis shows the flux in units of 10$^{-16}$ erg s$^{-1}$ cm$^{-2}$ \AA$^{-1}$ rescaled with the respect
to the blue panel by factors 2 and 0.5 for the yellow and red panel, respectively. }
\vspace{-0.2in}
\label{fig:basic}
\end{figure*}

Figure \ref{fig:basic} illustrates the general configuration of our observations performed with \vimos\
(\textbf{VI}sible \textbf{M}ulti-\textbf{O}bject \textbf{S}pectrograph) in the IFU mode on the VLT.
We have focused on the northwestern
rim, because it produces the brightest emission from a Balmer-dominated shock. As a first analysis of our data, we
divided the shock structure into four strips and bin up the data along each strip. The two-component
H$\alpha$ line is convincingly detected. To first order, the broad line width (FWHM) $W$ yields the shock velocity $v_s$,
whereas the level of energy equilibration between electrons and protons in the post-shock gas introduces small corrections \cite{cp88,ghava07}.
Because the broad line profile is a direct probe of the velocity distribution of the post-shock protons, and the narrow line profile traces pre-shock hydrogen atoms, the ratio of broad to narrow line intensities $I_{b}/I_{n}$ contains information on how energy is shared between the pre- and post-shock regions.
If no energy is shared, models which include the basic shock and atomic physics correctly translate the measured $W$ and
$I_b/I_n$ values into the inferred $v_s$ and $\beta$ values \cite{ghava02,v08,hm07,heng07}, where $\beta$ is the ratio
of electron to proton temperatures in the post-shock gas. Unusually low $I_b/I_n$ values ($\lesssim 0.7$) indicate that
suprathermal particles from the post-shock region are traveling upstream into the pre-shock gas and depositing energy via atomic
interactions (excitation, ionization and charge exchange), thus acting as precursors \cite{ghava00,sollerman03,lee07}. The loss of energy decreases the shock velocity and hence $W$.  The increased flux of particles into the pre-shock region leads to enhanced excitation of the hydrogen atoms via collisions and increases $I_n$.  The binned regions in Figure \ref{fig:basic} already hint at this phenomenon, because $I_b/I_n \approx 0.6$--0.7 in two of the strips (yellow and red). The broad line exhibits subtle deviations from a Maxwellian profile in the
line core, further supporting the presence of suprathermal particles in the shock.

\begin{figure*}[!t]
\begin{center}
\includegraphics[width=\textwidth,trim=0 7mm 0 0]{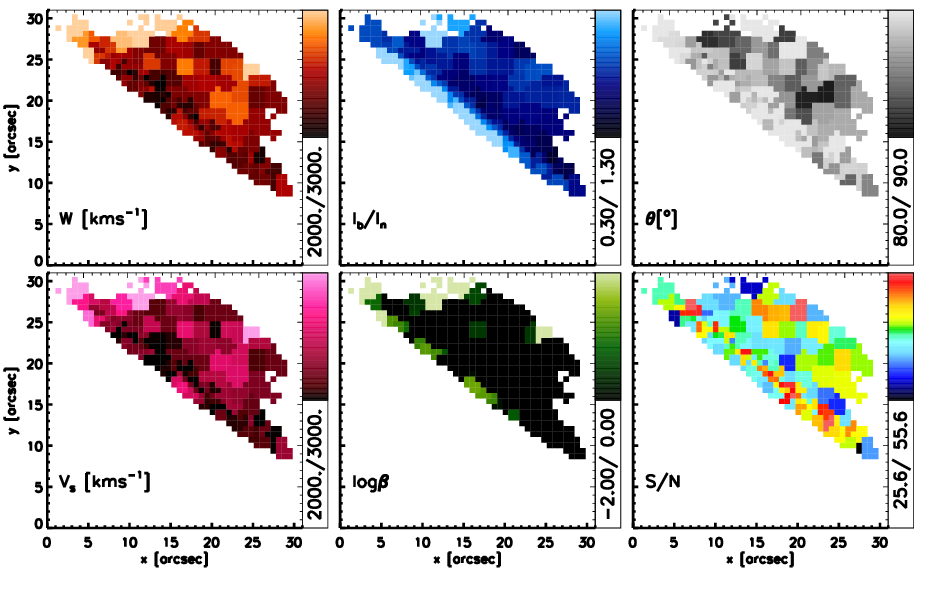}
\vspace{-0.2in}
\end{center}
\caption{Two-dimensional spatial-spectral maps of various properties associated with the shock front in the remnant of SN1006.
The \emph{bottom-right} panel shows the spatial Voronoi binning to reach a minimum signal-to-noise $S/N \approx 40$.
The maps in the \emph{top row} from left to the right: the broad line width $W$ (in \kms), the broad-to-narrow line
intensity ratio $I_b/I_n$, and the shock inclination angle $\theta$ (in degrees). Shock velocities $v_s$ and electron-to-proton
temperature ratios $\beta$ shown in the \emph{bottom-left} and \emph{bottom-middle} panels, respectively.}
\vspace{-0.2in}
\label{fig:voronoi}
\end{figure*}

We refined our approach by using the technique of Voronoi binning \cite{cc03}. We combined neighbouring spectra into spatial Voronoi bins until S/N of about 40 was reached (see lower-right panel of Figure~\ref{fig:voronoi}). We fitted double-Gaussians to the detected H$\alpha$ lines as shown in Section 2 of the
Supplementary Materials (hereafter SM) and derived $W$ and $I_b/I_n$ values,
thus producing the maps in the top-left and top-middle panel of Figure~\ref{fig:voronoi}. In addition to the
lack of complex density variations, we demonstrated that the viewing geometry of the shock is simple: by measuring the
velocity shift $\Delta V$ between the centroids of the narrow and broad H$\alpha$ line components, we derived the viewing
angle $\theta$ via the relation $\Delta V = 3 v_s \cos\theta / 4$. Our measurements indicate that $\theta \approx 80^\circ$--90$^\circ$.
Therefore, the shocks we observe are mostly edge-on as expected, but the rim seems slightly S-shaped along the line of sight
as suggested by \cite{ray07}.

Surprisingly, we detected significant spatial variations in $W$ and $I_{b}/I_{n}$ across length scales 10\arcsec$\sim$70$L_{\rm mfp}$, despite the
simple geometry of the shocks, suggesting that they arise from variations in the microphysics rather than density.
Within the bright rim the variations in $W$ are of order 10-20\%, significantly larger than the individual measurement
uncertainties (see Section 2 and Table S1 in SM). The density variations of 20-40\% required to explain the detected variations in $W$
are much larger than expected on these scales, and also incompatible with the
unchanged smoothness of the shock over two decades of imaging observations \cite{win03,ray07}.

\begin{figure}[!h]
\includegraphics[width=\columnwidth]{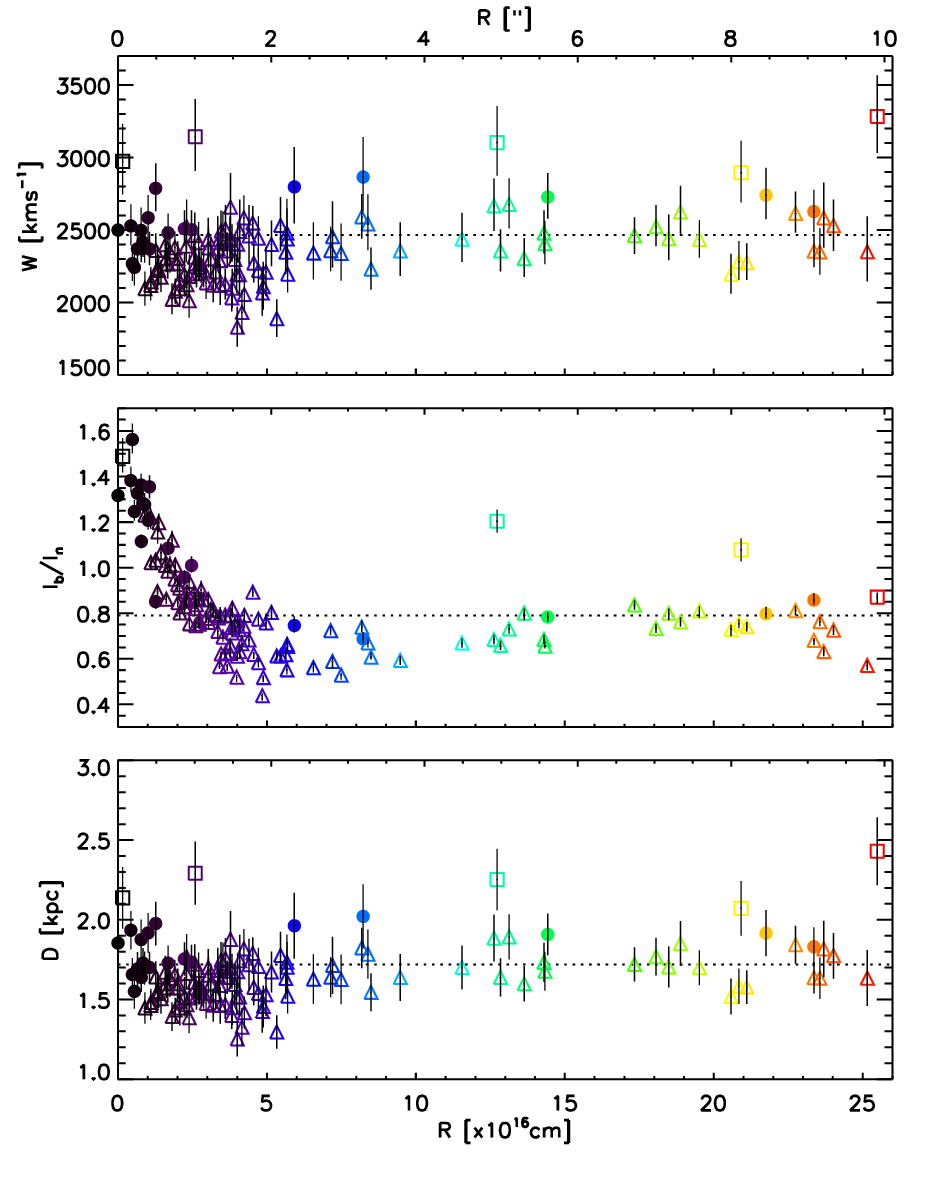}
\vspace{-0.4in}
\caption{The three panels show all measured (with error bars) broad line widths $W$, broad-to-narrow intensity ratios
$I_b/I_n$, and heliocentric distances $D$ from combining the proper motion measurement with the shock velocities. Data are
ordered in increasing distance from the inner rim (black/purple) to the outer rim (orange/red color), shown in arcsec
at the top and in units of $10^{16}$ cm at the bottom horizontal axis. The inner rim coincides with the inner edge of the
blue rim in the top-right panel in Figure~\ref{fig:basic}. The dashed horizontal lines indicate the measured
$W$=2465.76 \kms\ and $I_b/I_n$=0.79 values from collapsing all spectra of the pixels on the shock front, and from there
the inferred $D$=1.72 kpc. Data points marked with a filled circle are
those for which a valid model solution is found. The triangles (squares) indicate that due to low
(high) $I_b/I_n$ values the models hit the lower (upper) boundary limit in $\beta$ of 0.01
(1.0).}
\vspace{-0.2in}
\label{fig:spatial}
\end{figure}

The low $I_b/I_n$ values are found at all distances from the inner rim, reaching as low as 0.4 at some locations (Figure~\ref{fig:spatial}). We used the models without non-thermal physics to infer the values of $v_s$ and $\beta$ \cite{v08}. Most (about 85\%) of the binned data are not accounted for by
the model because of the low $I_b/I_n$ values, thus motivating the need for models which include suprathermal
particles and cosmic rays \cite{morcr12}. Combining the derived shock velocities with the proper motion measurement from \cite{win03},
we obtained estimates of the heliocentric distance to SN\,1006. Clearly, there has to be a unique distance, likely given by the upper
points around $\sim$ 2 kpc (see also Section 3 in SM). Most points, however, are underestimating the distance due to loss of energy from the broad
line component resulting in too low inferred shock velocities.

These low values and variations of $W$ and $I_b/I_n$, as well as the potential non-Gaussian contributions to the broad H$\alpha$
lines, demand an explanation.  We examined three possibilities. The first is that ``broad neutrals" are acting as a precursor
\cite{lr96,morbn12}. These are secondary populations of ``hot" hydrogen atoms which are produced when post-shock protons capture an electron
from a pre-shock hydrogen atom---the subsequent excitation of these broad neutrals produces the broad H$\alpha$ line component
detected in our observations \cite{cr78,bl79,ckr80,hm07}.
Broad neutrals will warm the pre-shock gas and produce a third H$\alpha$ line component of intermediate width, which may
account for the narrow-plus-broad-line double-Gaussian being unable to fit the data within the measured uncertainties (i.e., not
reaching a reduced $\chi^2$ of unity). The strongest argument against broad neutrals is that they can only act
as a precursor over an atomic mean free path, which is at odds with our observational result that the variations and low
values of $W$ and $I_b/I_n$ extend over distances $\gg L_{\rm mfp}$.

The second possibility is that pre-shock hydrogen atoms may cross the shock front, eventually become ionized and become protons
gyrating along an ambient magnetic field line. It has been shown that these ``pick-up protons" settle into a bi-spherical distribution
which introduces a non-Gaussian contribution to the broad H$\alpha$ line core \cite{ray08}. This explanation requires the
magnetic field to be ordered on small length scales ($\ll L_{\rm mfp}$), whereas turbulent magnetic fields will result in a broad
H$\alpha$ line that is approximately Gaussian.

Both explanations are disfavored at low pre-shock neutral fractions, which is the situation in SN\,1006 with
a pre-shock neutral fraction of about 0.1 \cite{ghava02}. This is because both broad neutrals and pick-up ions require the
pre-shock gas to contain a substantial population of hydrogen atoms (relative to electrons and protons) in order to initiate
the process. In the limit of a fully ionized pre-shock gas, no broad neutrals or pick-up ions may be produced.

The explanation we favor is that the post-shock proton population includes a non-thermal sub-population of protons---suprathermal
protons (that are not pick-up protons). Such an explanation requires no assumption on the magnetic field geometry or pre-shock
neutral fraction. The lack of
non-thermal X-ray and TeV $\gamma$-ray emission in the northwestern rim of the remnant indicates that if there is a CR precursor,
the injection of electrons is little and the particle energies are low ($<$ 1MeV), but CR acceleration can still happen.

Our pilot project demonstrates the feasibility of using integral-field spectroscopy
to observe and study the microphysics of high-velocity shocks around supernova
remnants. The resulting high spatial resolution mapping of the Balmer dominated
shocks in the northwestern rim of SN\,1006 suggests the presence of suprathermal
protons of energies 10–-100 keV which can seed high-energy cosmic rays.

\small
\textbf{Acknowledgements:}
This work was supported by the IMPRS for Astronomy \& Cosmic Physics at the University of Heidelberg.
The paper is based on observations made with European Southern Observatory Telescopes at the La Silla Paranal
Observatory under programme ID 085.D-0983.

\newpage
\renewcommand{\thefigure}{S1}
\renewcommand{\thetable}{S\arabic{table}}
{\Large\textbf{Supplementary Materials}}\\
\\
\\
\indent
This Supplemental Material consists of three parts. The first part describes our
intergral-field spectroscopic observations and data reduction. This is followed by an
assessment of the accuracy in fitting the H$\alpha$ emission line profile to extract
the shock properties. Finally, we discuss estimates of the heliocentric distance to
SN\,1006. Figures S1 through S3 support the H$\alpha$-profile
fitting, with resulting parameters and uncertainties as well as shock properties
given in Table S1.

\section{\textbf{Integral-field Spectroscopic Observations and Data Reduction}}

Our observations of the northwestern rim of SN\,1006, were performed in queue mode on the nights of 2010 April 6,
11, 12, 13, and May 10, using the \vimos\ (\textbf{VI}sible \textbf{M}ulti-\textbf{O}bject \textbf{S}pectrograph)
in the IFU (integral field unit) mode on the VLT.  The IFU was placed approximately 15\arcmin\ away from the remnant's center, at coordinates $\alpha_{J2000}=15^{h}02^{m}13^{s}.5$, $\delta_{J2000}=-41^{^\circ}45^{'}22^{''}$.  The overall exposure time of 6 hours accumulated in 10 exposures, all taken at air mass $\lesssim$ 1.60 and seeing $\lesssim$ 1\arcsec.60, enabled us to collect enough photons to reach the required signal-to-noise ratios (S/N).  A dither pattern with small offsets between the individual exposures was applied in order to account for dead fibers.  We have used a spatial scale of 0\arcsec.67 per pixel in combination with the HR-Orange grism providing a field-of-view of 27\arcsec$\times$27\arcsec as well as a resolution of R$\approx$2650 within the wavelength range of $5250\,\AA\lesssim\lambda\lesssim7400\,\AA$.

The data has been reduced with our own dedicated data reduction pipeline which was developed in particular to deal
with the rather complex \vimos\ instrument in IFU mode. It is based on several individual scripts written in the
software language \texttt{python}. \vimos\ is made of four identical optical quadrants, each representing a
completely independent spectrograph.  Each quadrant is at first reduced separately, including the following
standard procedures: bias subtraction, straylight correction, interactive fibre-identification and -tracing, spectra
extraction and wavelength calibration using the arclamp frames. For the extraction of the fluxes the ``optimal
extraction algorithm" was applied \cite{sharp10}, which was designed to deal with dense packed IFUs to correct for the
cross talk between adjacent fibres. The associated continuum lamp frames were used to construct corresponding
fibre-flatfields which were used to normalise the wavelength-dependent throughput of each fibre. Additionally,
\vimos\ also suffers from substantial distortions on the raw frames due to instrumental flexure. Thus, in order to
guarantee the best possible extraction results the fiber traces were smoothed and corrected for shifts between science
and calibration frames. This included a 2nd order correction of the absolute positions of the traces as well as an
adjustment of the wavelength solution by using the available strong night-sky emission lines within the observed
wavelength-range.  The raw frames were cleaned for cosmic rays using the newly developed software package \texttt{PyCosmics} \cite{hus12},
which represents an optimized algorithm to detect their artifacts in IFU raw-data.

After the basic reduction steps the four individual science frames for each observing block were combined and rearranged into three dimensional datacubes, using the dedicated \vimos\ lookup-tables. Flux calibration was achieved by reduction and extraction of photometric standard stars which were observed using the same setup as the science data. As the \vimosifu\ does not have ``sky-dedicated" fibers, it is normal practice to do separate offset ``sky exposures", but in this case the sky background could be reliably measured from the ``empty" downward triangular part in the field-of-view (see top-right panel of Figure~1).  The last step involved the correction of the barycentric Doppler shift, i.e., the offset in velocity space caused by the movement of the Earth around the Sun. The observations were accomplished within a time-window of approximately one month, causing a velocity shift of about 20 km s$^{-1}$ which corresponds roughly to one pixel on the detector. In order to readjust this offset we measured the center of the narrow H$\alpha$ emission line and shifted it to the laboratory wavelength of 6562.8\,\AA.

The H$\alpha$ line is found in the middle of the HR-Orange grism wavelength range of 5250--7400\,\AA. This spectral range is
broad enough to both easily fit the broad component of the H$\alpha$ line with width of the order of 50\,\AA, and to still
leave plenty of spectral range for an accurate determination of the continuum.  Finally, we used the precise header information of the telescope-pointings to compute the spatial-offsets between each observation and combined all ten individual datacubes within one final cube, containing in total 2162 spectra.
The instrumental resolution of about 110\,\kms\ (FWHM) around the wavelength of H$\alpha$ is more than sufficient to measure the width of the broad line of 2000--3000\,kms$^{-1}$, as well as deviations from a Gaussian profile. The width of the narrow line, expected to be about 20 km s$^{-1}$, cannot be resolved, but the narrow line intensity can be accurately measured, especially given that the narrow line has a significantly higher S/N than the broad line.

\vspace{0.1in}
\section{\textbf{Accuracy of the H$\alpha$-line fitting}}

We fitted the H$\alpha$-line with two Gaussians, one for the narrow component and the other for the broad component, after
convolving with a Gaussian instrumental profile.
Hint on the non-Gaussianity around the line core of the broad component that we detected is a supporting, but not the main piece
of evidence of suprathermal protons presence.

\begin{figure*}[ht]
\vspace{-4in}
\includegraphics[width=0.5\textwidth,angle=-90]{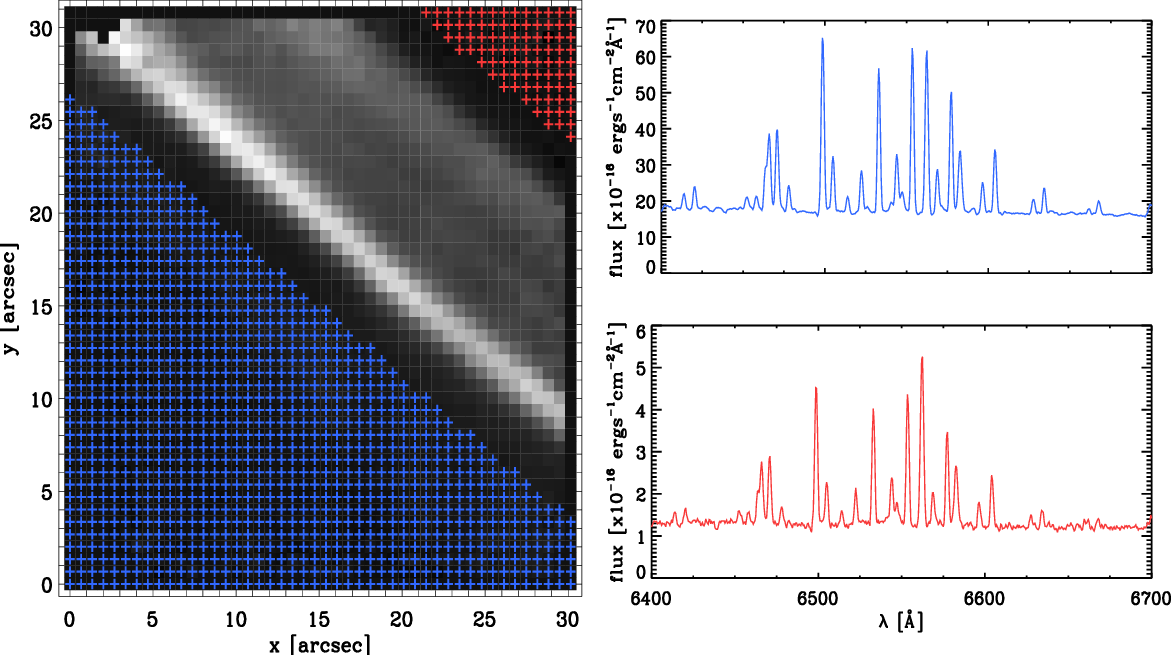}
\caption{The \emph{left} panel shows the reduced data cube collapsed in wavelength around the H$\alpha$-line with blue
and red crosses indicating the pixels for which the combined non-sky-subtracted spectra are shown in the \emph{right} panels.
Comparing the continuum emission across wavelength, we confirm that the emission profile is the same in the upstream and
downstream region of the shock. }
\label{fig:continuumprofile}
\end{figure*}

In order to investigate possible non-physical contributions to the H$\alpha$-line profile, we compared the continuum emission profile
across wavelength in the region downstream and upstream of the shock. As shown in Figure \ref{fig:continuumprofile}, we combined spectra in the downstream and
upstream regions. The continuum has the same profile in these two regions, and does not
affect the H$\alpha$-line profile. In these regions we did, however, detect diffuse narrow-line H$\alpha$ emission, with
surface brightness, 10-30 times lower compared to the emission in the four regions in Figure 1.

Next, to verify that the instrumental profile is Gaussian we fitted sky lines that went through
the same data reduction steps (except for sky subtraction). In Figure \ref{fig:instrprofile}, we show five sky lines
around H$\alpha$ line fitted with a Gaussian. The fits together with the differences between the observed and
fitted sky lines show that the instrumental profile is indistinguishable from Gaussian.

We have also checked if the residuals might be coming
from the intrinsic variations among the individual spectra that went into the combined spectra. The offsets between velocity
centroids are much smaller than the "extra-core" size, and thus rule out this possibility. Also, the individual bins have
very similar narrow line widths that are always unresolved, but much smaller than the "extra-core" size.

Finally, the uncertainties in measuring the broad line width $W$ and broad-to-narrow intensity ratio $I_b/I_n$ were
determined in various independent ways, taking into account the varying noise per pixel in wavelength as given by the derived error spectrum per pixel. Two of the methods were very robust and led to consistent error estimates even in case of spectra with signal-to-noise significantly lower than our target $S/N \approx 40$. In the first method, we created 1000 mock spectra per Voronoi bin through Monte Carlo
sampling of the observed (combined) spectrum, and fitted each mock spectrum with a double Gaussian. The resulting
distributions of values in $W$ and $I_b/I_n$ shows that both quantities are uncorrelated and their means are the
same as from the fit to the observed spectrum. Their standard deviations, however, are slightly larger than those
from the second method in which we varied\footnote{Keeping the total intensity $I_n + I_b$, narrow line width and
centroids of the narrow and broad component fixed to their best-fit values.} $W$ and $I_b/I_n$ on a grid around
their best-fit values until $\chi^2$ changed by an amount corresponding to the 68\% confidence level given the
degrees of freedom. This is expected, as the first method, apart from the added noise from the error spectrum,
also includes noise already inherent in the observed spectrum, increasing the reduced $\chi^2$ values of the mock
spectra fits above unity. Re-scaling the reduced $\chi^2$ values to unity leads to uncertainties that are fully
consistent with those inferred from the second method. The double-Gaussian fits to the H$\alpha$-line profile in all
133 sky positions are shown in Figure \ref{fig:profiles}. The resulting best fit parameters and uncertainties are given in Table S1,
along with shock properties inferred using van Adelsberg models.

\vspace{0.1in}
\section{\textbf{Heliocentric distance to SN\,1006}}

When converting from angular to physical length scales, 1\arcsec $\approx$ 2.58$\times$10$^{16}$ cm, we adopted a heliocentric
distance of 1.72 kpc, derived as follows. We first summed up all pixels on the shock front, and extracted
the width of the broad component $W$ and intensity ratio $I_{b}/I_{n}$ by fitting the H$\alpha$ line with double-Gaussians.
The resulting broad line width and broad-to-narrow intensity ratio are $W$ = 2466 \kms\ and $I_{b}/I_{n}$ = 0.79. Using the van Adelsberg et al. model,
this yields a shock velocity of $v_{s}$=2288 \kms. Combining this shock velocity with the proper motion measurements of 280 \masyr\ [9]
yields a heliocentric distance of 1.72 kpc.\\
\indent Previously inferred
distances to SN\,1006 from the H$\alpha$ observations in the northwestern rim, 2.18 kpc [9] and 1.6 kpc [19],
were based on long-slit spectroscopic observations from [13].
The discrepancy in the inferred distances comes from different shock models applied.
The latter one uses the same van Adelsberg et al. model, yielding a lower shock velocity for the same $W$ = 2290 \kms and
$I_b/I_n$ = 0.84, primarly due to the broad neutral velocities contribution to the relative
speeds in fast neutral-ion interactions.\\
\indent Analysis of the optical spectrum of the Schweizer-Middleditch star \cite{bur00}
which lies almost at the same line-of-sight as the center of SN\,1006, sets the upper limit on the distance of SN\,1006 at 2.1 kpc.
The observed ejecta expansion at 7026 \kms\ \cite{ham07}, and the requirement that
this material lie within the remnant places a lower limit of 1.6 kpc.
We actually expect the distance to be closer to $\sim$ 2 kpc to explain the upper envelope
of the points in Figure 3, but a model that properly takes into account the non-thermal
physics is needed before we can robustly measure an intrinsic shock velocity and correspondingly
obtain a secure estimate of the heliocentric distance.


\renewcommand{\thefigure}{S2}
\begin{figure*}[ht]
\begin{center}
\includegraphics[width=0.9\textwidth]{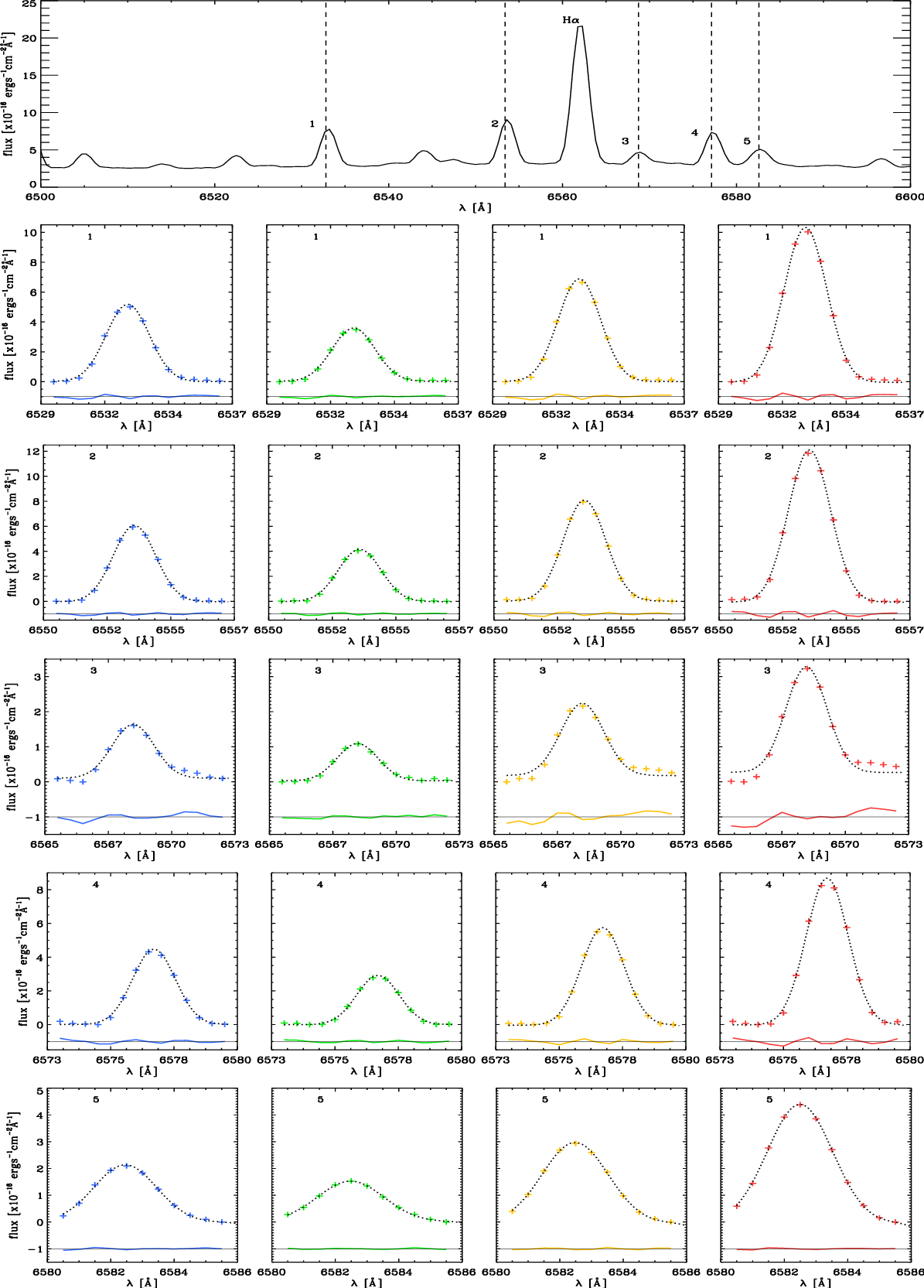}
\caption{The \emph{top} panel shows a non-sky-subtracted spectrum with five sky lines around H$\alpha$, indicated by vertical dashed lines and numbered.
Next, each \emph{row} shows one sky line fitted with a Gaussian for the four different regions consistent with the ones in Figure 1. The fits along with
the differences between observed and fitted sky lines (with an offset downward for clarity) show that the instrumental profile is Gaussian.
 }

\label{fig:instrprofile}
\end{center}
\end{figure*}

\renewcommand{\thefigure}{S3}
\begin{figure*}[!ht]
\vspace{0.15in}
\includegraphics[width=\textwidth]{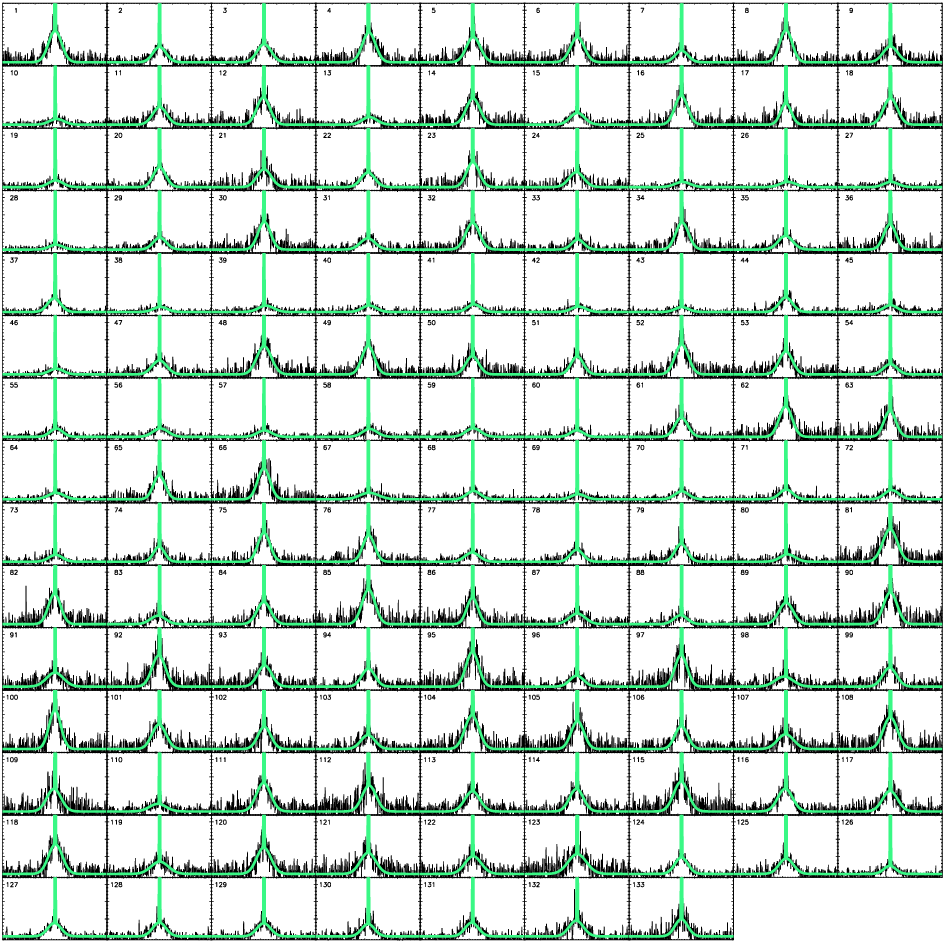}
\caption{Montage of the H$\alpha$ line profiles in each of the 133 Voronoi bins. The horizontal axis represents wavelengths and vertical axis is in the relative units. Over-plotted are the double-Gaussian fits to the narrow and broad line components.  No smoothing has been applied to the data.}
\label{fig:profiles}
\end{figure*}

\onecolumn
\renewcommand{\thetable}{S\arabic{table}}
\tablecaption{Properties of the shock front in the remnant of SN\,1006 for 133 spatially binned locations.  Columns 1--5: number of the (Voronoi) bin, $x$ and $y$ coordinates of the bin centroid, number of combined pixels, and signal-to-noise ratio. Columns 6--10: measured values of the broad component $W$ and the broad-to-narrow line intensity ratio $I_b/I_n$ (both with estimated uncertainties based on $\chi^2$ 68\% confidence levels), shock inclination angle $i$ (with typical uncertainty of 1.6$^{\circ}$ coming from a typical error of 47 \kms in $\Delta$V), and the reduced $\chi^2$ of the double-Gaussian fit to the observed H$\alpha$-line. Columns 10--11: shock velocity $v_s$ and electron-to-proton ratio $\beta$ (in base-10 logarithm) of the best-fit shock model; a long dash indicates when the shock model hits the adopted boundary limits of $\log\beta=(-2,0)$, likely because CR physics is missing.}
\label{tab:sn1006_bins}
{\small
\tablehead{\hline
Bin & x& y& Spx & $S/N$
& $W_\mathrm{obs}$ 
& $(I_b/I_n)_\mathrm{obs}$ 
& $i_\mathrm{obs}$
&  red.\ $\chi^{2}$ & $v_{s}$ & $\log \beta$ \\
& \arcsec & \arcsec & & &\kms & & \dgr & & \kms & \\
(1) & (2) & (3) & (4) & (5) & (6) & (7) & (8) & (9) & (10) & (11) \\
\hline}
\tabletail{\hline}
\begin{xtabular}{|c|c|c|c|c|c|c|c|c|c|c|}
  1&19.43&16.08&  1&    52.68& 2268 (+110/ -94)&     0.93 (+0.03/-0.03)& 89.3&  0.91& 2088&------\\\hline

  2&19.43&15.38&  3&    36.57& 2242 (+134/-126)&     1.25 (+0.05/-0.04)& 89.7&  0.78& 2058&-1.86\\\hline

  3&18.76&16.03&  3&    50.67& 2417 (+110/-102)&     1.28 (+0.03/-0.04)& 89.1&  0.91& 2294&-0.95\\\hline

  4&20.10&15.41&  1&    44.70& 2267 (+110/-102)&     0.98 (+0.03/-0.03)& 86.9&  0.82& 2087&------\\\hline

  5&20.10&16.08&  1&    48.21& 2376 (+141/-134)&     0.76 (+0.02/-0.03)& 87.2&  0.81& 2194&------\\\hline

  6&19.43&16.75&  1&    50.81& 2370 (+141/-126)&     0.78 (+0.02/-0.03)& 89.2&  0.78& 2189&------\\\hline

  7&20.24&16.79&  3&    44.35& 2219 (+157/-141)&     0.58 (+0.02/-0.02)& 89.5&  0.81& 2041&------\\\hline

  8&18.76&16.75&  1&    45.85& 2124 (+ 94/ -94)&     0.89 (+0.03/-0.03)& 89.5&  0.75& 1948&------\\\hline

  9&20.77&16.08&  1&    36.70& 2490 (+220/-188)&     0.63 (+0.03/-0.03)& 89.9&  0.72& 2313&------\\\hline

 10&20.93&18.06&  8&    38.61& 2542 (+204/-181)&     0.67 (+0.02/-0.02)& 82.7&  1.28& 2368&------\\\hline

 11&18.85&17.42&  2&    51.71& 2406 (+134/-126)&     0.71 (+0.02/-0.02)& 89.3&  0.92& 2226&------\\\hline

 12&20.77&15.41&  1&    43.55& 2266 (+118/-110)&     0.77 (+0.02/-0.03)& 86.7&  0.81& 2086&------\\\hline

 13&18.90&18.18&  4&    36.01& 2531 (+196/-173)&     0.61 (+0.02/-0.02)& 84.6&  0.86& 2357&------\\\hline

 14&18.09&16.75&  1&    44.23& 2360 (+118/-110)&     1.03 (+0.03/-0.04)& 88.9&  0.85& 2179&------\\\hline

 15&17.76&16.39&  4&    38.21& 2499 (+134/-118)&     1.32 (+0.05/-0.04)& 89.7&  0.89& 2461&-0.81\\\hline

 16&18.09&17.42&  1&    51.08& 2199 (+102/ -94)&     0.84 (+0.03/-0.02)& 88.1&  0.83& 2021&------\\\hline

 17&18.09&18.09&  1&    38.59& 1829 (+141/-134)&     0.61 (+0.02/-0.02)& 89.5&  0.97& 1663&------\\\hline

 18&17.42&17.42&  1&    43.81& 2399 (+126/-118)&     1.01 (+0.04/-0.03)& 89.5&  0.79& 2218&------\\\hline

 19&21.60&16.22&  5&    35.75& 2479 (+236/-204)&     0.55 (+0.02/-0.02)& 87.4&  0.93& 2302&------\\\hline

 20&20.62&14.74&  3&    46.98& 2143 (+102/ -94)&     1.02 (+0.02/-0.03)& 89.8&  0.79& 1966&------\\\hline

 21&21.44&15.41&  1&    38.59& 2432 (+165/-157)&     0.66 (+0.03/-0.02)& 89.6&  0.89& 2252&------\\\hline

 22&21.36&14.01&  4&    42.77& 2372 (+110/ -94)&     1.12 (+0.03/-0.03)& 87.6&  0.87& 2175&-1.31\\\hline

 23&21.44&14.74&  1&    43.37& 2181 (+149/-126)&     0.75 (+0.03/-0.02)& 89.4&  0.83& 2003&------\\\hline

 24&22.11&14.82&  2&    44.82& 2507 (+165/-149)&     0.63 (+0.02/-0.02)& 86.3&  0.82& 2331&------\\\hline

 25&22.76&18.73& 14&    42.71& 2667 (+188/-173)&     0.68 (+0.02/-0.02)& 85.8&  1.40& 2502&------\\\hline

 26&21.09&20.26& 12&    41.81& 2676 (+181/-165)&     0.73 (+0.02/-0.02)& 81.3&  1.25& 2512&------\\\hline

 27&23.05&15.86&  8&    33.32& 2454 (+243/-212)&     0.59 (+0.02/-0.03)& 86.8&  1.01& 2275&------\\\hline

 28&18.79&20.23&  8&    31.57& 2356 (+196/-173)&     0.59 (+0.02/-0.02)& 85.3&  1.10& 2175&------\\\hline

 29&17.54&18.78&  3&    46.08& 2272 (+134/-134)&     0.62 (+0.01/-0.02)& 88.8&  0.95& 2091&------\\\hline

 30&17.42&18.09&  1&    48.11& 2135 (+118/-110)&     0.78 (+0.03/-0.02)& 88.4&  0.82& 1959&------\\\hline

 31&16.72&17.39&  4&    35.13& 2528 (+149/-134)&     1.38 (+0.06/-0.05)& 87.2&  0.86& 2567&-0.71\\\hline

 32&16.75&18.09&  1&    47.90& 2379 (+110/-102)&     0.95 (+0.03/-0.02)& 89.4&  0.80& 2198&------\\\hline

 33&16.86&19.45&  3&    39.35& 2108 (+173/-157)&     0.52 (+0.01/-0.02)& 86.1&  1.06& 1933&------\\\hline

 34&16.75&18.76&  1&    51.06& 2211 (+110/-102)&     0.79 (+0.02/-0.02)& 87.2&  0.84& 2032&------\\\hline

 35&22.80&14.22&  3&    49.91& 2493 (+141/-134)&     0.62 (+0.01/-0.02)& 87.6&  0.87& 2316&------\\\hline

 36&22.11&14.07&  1&    37.59& 2095 (+141/-126)&     0.80 (+0.03/-0.03)& 87.2&  0.79& 1920&------\\\hline

 37&22.61&13.32&  5&    46.21& 2228 (+102/ -94)&     0.89 (+0.02/-0.02)& 84.5&  0.98& 2050&------\\\hline

 38&25.73&17.29& 22&    42.98& 2406 (+157/-141)&     0.66 (+0.02/-0.01)& 86.5&  1.49& 2226&------\\\hline

 39&23.52&21.21& 11&    41.18& 2624 (+181/-165)&     0.76 (+0.02/-0.02)& 80.4&  1.28& 2456&------\\\hline

 40&21.88&22.28&  9&    36.97& 2441 (+165/-149)&     0.80 (+0.02/-0.03)& 85.7&  1.20& 2262&------\\\hline

 41&25.39&20.75& 12&    45.41& 2281 (+141/-126)&     0.75 (+0.02/-0.02)& 83.6&  1.24& 2101&------\\\hline

 42&19.31&22.20&  9&    34.71& 2478 (+157/-141)&     0.68 (+0.02/-0.02)& 83.0&  1.30& 2300&------\\\hline

 43&24.61&14.80&  9&    31.54& 2339 (+204/-188)&     0.53 (+0.01/-0.02)& 88.4&  1.00& 2158&------\\\hline

 44&23.49&13.67&  3&    49.25& 2289 (+126/-118)&     0.57 (+0.02/-0.01)& 88.5&  0.93& 2109&------\\\hline

 45&16.63&20.46&  5&    38.63& 2341 (+212/-181)&     0.56 (+0.02/-0.02)& 89.7&  1.06& 2160&------\\\hline

 46&17.48&22.25&  8&    36.44& 2437 (+181/-157)&     0.67 (+0.02/-0.02)& 84.5&  1.13& 2258&------\\\hline

 47&16.03&18.09&  3&    36.60& 2498 (+157/-141)&     1.36 (+0.05/-0.05)& 89.3&  0.86& 2490&-0.76\\\hline

 48&16.08&18.76&  1&    48.57& 2509 (+126/-118)&     0.96 (+0.03/-0.03)& 89.5&  0.80& 2328&-1.90\\\hline

 49&16.08&19.43&  1&    46.59& 2124 (+110/-102)&     0.79 (+0.02/-0.03)& 89.1&  0.67& 1949&------\\\hline

 50&16.08&20.10&  1&    33.76& 2210 (+173/-165)&     0.76 (+0.04/-0.03)& 87.5&  0.83& 2031&------\\\hline

 51&15.36&18.75&  3&    38.40& 2117 (+110/-102)&     1.22 (+0.04/-0.05)& 90.0&  0.93& 1941&------\\\hline

 52&15.41&19.43&  1&    47.63& 2313 (+126/-118)&     0.89 (+0.03/-0.02)& 89.5&  0.87& 2133&------\\\hline

 53&15.41&20.10&  1&    39.34& 2297 (+165/-149)&     0.75 (+0.03/-0.03)& 89.0&  0.63& 2117&------\\\hline

 54&15.19&21.11&  4&    41.04& 2347 (+173/-149)&     0.62 (+0.02/-0.02)& 89.2&  0.92& 2166&------\\\hline

 55&20.45&24.40&  6&    39.26& 2194 (+141/-134)&     0.73 (+0.02/-0.03)& 83.2&  1.06& 2016&------\\\hline

 56&21.87&24.37&  7&    44.35& 2615 (+149/-134)&     0.81 (+0.02/-0.02)& 85.6&  1.16& 2446&------\\\hline

 57&18.87&24.37&  9&    41.68& 2522 (+149/-134)&     0.73 (+0.02/-0.02)& 87.5&  1.37& 2348&------\\\hline

 58&25.24&22.42&  8&    39.97& 2529 (+181/-173)&     0.73 (+0.02/-0.02)& 83.5&  1.00& 2355&------\\\hline

 59&23.15&23.69&  7&    43.80& 2627 (+149/-141)&     0.86 (+0.03/-0.02)& 87.7&  0.97& 2430&-1.57\\\hline

 60&15.20&22.50&  8&    36.31& 2230 (+149/-141)&     0.61 (+0.02/-0.01)& 88.0&  1.25& 2051&------\\\hline

 61&14.65&19.43&  3&    38.76& 2220 (+110/-102)&     1.20 (+0.04/-0.04)& 88.6&  0.91& 2042&------\\\hline

 62&14.74&20.10&  1&    44.37& 2225 (+118/-118)&     0.86 (+0.03/-0.03)& 88.0&  0.75& 2046&------\\\hline

 63&14.74&20.77&  1&    36.95& 2055 (+126/-118)&     0.79 (+0.03/-0.03)& 87.0&  0.78& 1881&------\\\hline

 64&16.76&24.21&  9&    44.92& 2727 (+165/-149)&     0.78 (+0.02/-0.02)& 86.1&  1.20& 2533&-1.57\\\hline

 65&14.07&20.10&  2&    35.87& 2022 (+110/-102)&     1.12 (+0.04/-0.05)& 88.2&  0.80& 1849&------\\\hline

 66&14.07&20.77&  1&    42.29& 2120 (+134/-118)&     0.81 (+0.03/-0.03)& 89.0&  0.84& 1944&------\\\hline

 67&24.64&23.59&  7&    38.71& 3284 (+283/-251)&     0.87 (+0.02/-0.03)& 86.8&  0.92& 3225&------\\\hline

 68&27.50&20.46& 16&    41.68& 2348 (+165/-157)&     0.76 (+0.02/-0.02)& 86.1&  1.13& 2168&------\\\hline

 69&21.78&25.62&  9&    34.81& 2350 (+243/-204)&     0.57 (+0.02/-0.02)& 84.2&  1.12& 2169&------\\\hline

 70&20.09&26.04& 11&    53.82& 2354 (+118/-110)&     0.68 (+0.01/-0.02)& 88.9&  1.36& 2173&------\\\hline

 71&18.43&26.20&  6&    46.42& 2275 (+134/-118)&     0.74 (+0.02/-0.02)& 89.1&  1.26& 2094&------\\\hline

 72&17.12&26.42&  6&    46.57& 2435 (+134/-126)&     0.81 (+0.02/-0.03)& 83.7&  1.09& 2256&------\\\hline

 73&14.81&24.92&  9&    37.19& 2356 (+149/-141)&     0.66 (+0.02/-0.02)& 87.7&  1.27& 2175&------\\\hline

 74&13.66&22.13&  3&    37.38& 1889 (+134/-126)&     0.61 (+0.02/-0.02)& 86.9&  0.96& 1720&------\\\hline

 75&13.40&20.66&  2&    45.33& 2078 (+ 86/ -86)&     1.01 (+0.03/-0.03)& 89.7&  0.88& 1903&------\\\hline

 76&13.59&21.44&  2&    51.06& 2032 (+102/ -94)&     0.74 (+0.03/-0.02)& 88.2&  0.84& 1859&------\\\hline

 77&15.49&26.61&  9&    47.53& 2462 (+126/-126)&     0.84 (+0.02/-0.02)& 86.8&  1.09& 2284&------\\\hline

 78&12.91&22.88&  5&    39.35& 2196 (+134/-126)&     0.65 (+0.02/-0.02)& 87.2&  1.06& 2018&------\\\hline

 79&12.67&20.72&  4&    35.90& 2096 (+118/-118)&     1.23 (+0.05/-0.04)& 86.8&  1.07& 1921&------\\\hline

 80&18.31&27.56&  8&    35.97& 2583 (+243/-204)&     0.63 (+0.02/-0.02)& 82.7&  1.14& 2412&------\\\hline

 81&12.73&21.44&  1&    44.55& 2502 (+134/-126)&     1.01 (+0.04/-0.03)& 88.3&  0.86& 2301&-1.31\\\hline

 82&12.73&22.11&  1&    39.36& 2099 (+134/-134)&     0.82 (+0.03/-0.04)& 86.3&  0.78& 1924&------\\\hline

 83&12.38&24.51&  8&    40.66& 2591 (+157/-149)&     0.74 (+0.03/-0.02)& 82.6&  1.13& 2420&------\\\hline

 84&11.83&21.44&  3&    41.26& 2369 (+118/-110)&     1.35 (+0.05/-0.05)& 89.1&  0.82& 2258&-0.91\\\hline

 85&12.06&22.11&  1&    41.05& 2259 (+134/-126)&     0.90 (+0.04/-0.04)& 89.1&  0.74& 2080&------\\\hline

 86&12.06&22.78&  1&    33.65& 1931 (+157/-141)&     0.67 (+0.04/-0.03)& 87.6&  0.82& 1761&------\\\hline

 87&16.59&27.92& 10&    42.13& 2741 (+188/-165)&     0.80 (+0.03/-0.02)& 89.4&  0.95& 2543&-1.45\\\hline

 88&13.41&26.38& 11&    39.36& 2302 (+141/-126)&     0.80 (+0.02/-0.03)& 84.4&  1.24& 2122&------\\\hline

 89&11.54&23.45&  2&    44.96& 2444 (+149/-134)&     0.77 (+0.03/-0.03)& 89.6&  0.77& 2265&------\\\hline

 90&11.39&22.11&  1&    38.93& 2348 (+141/-141)&     1.01 (+0.04/-0.04)& 89.7&  0.88& 2167&------\\\hline

 91&14.48&29.13& 12&    29.53& 2895 (+220/-204)&     1.08 (+0.05/-0.05)& 89.0&  1.11& 2749&------\\\hline

 92&11.39&22.78&  1&    42.26& 2302 (+149/-141)&     0.81 (+0.03/-0.03)& 85.4&  0.93& 2122&------\\\hline

 93&10.94&24.12&  2&    42.12& 2400 (+165/-149)&     0.80 (+0.03/-0.03)& 86.9&  0.80& 2220&------\\\hline

 94&10.62&22.08&  4&    35.85& 2268 (+126/-118)&     1.56 (+0.07/-0.06)& 90.0&  0.92& 2199&-0.80\\\hline

 95&10.72&22.78&  1&    37.78& 2151 (+134/-118)&     0.91 (+0.04/-0.04)& 87.2&  0.76& 1975&------\\\hline

 96&10.99&25.06&  6&    37.39& 2357 (+165/-141)&     0.72 (+0.02/-0.02)& 87.6&  1.15& 2176&------\\\hline

 97&10.72&23.45&  1&    40.19& 2117 (+149/-134)&     0.79 (+0.04/-0.03)& 86.8&  0.81& 1941&------\\\hline

 98&11.35&27.51& 15&    34.79& 3103 (+251/-228)&     1.20 (+0.05/-0.05)& 82.9&  1.19& 2990&------\\\hline

 99& 9.97&22.78&  3&    37.83& 2392 (+126/-118)&     1.28 (+0.05/-0.05)& 86.1&  0.91& 2253&-0.98\\\hline

100&10.05&23.45&  1&    40.63& 2013 (+118/-118)&     0.94 (+0.03/-0.04)& 88.6&  0.80& 1840&------\\\hline

101&10.05&24.25&  2&    49.73& 2402 (+134/-126)&     0.77 (+0.02/-0.03)& 87.2&  0.95& 2222&------\\\hline

102& 9.38&23.45&  2&    35.76& 2237 (+157/-149)&     1.16 (+0.05/-0.05)& 87.9&  0.94& 2058&------\\\hline

103& 9.50&25.48&  3&    37.77& 2438 (+212/-188)&     0.66 (+0.03/-0.02)& 87.8&  0.90& 2259&------\\\hline

104& 9.38&24.12&  1&    43.11& 2412 (+141/-126)&     0.86 (+0.03/-0.03)& 85.6&  0.70& 2232&------\\\hline

105& 9.38&24.79&  1&    38.80& 2191 (+165/-141)&     0.73 (+0.04/-0.03)& 88.2&  0.85& 2013&------\\\hline

106& 8.70&24.02&  3&    43.51& 2175 (+118/-102)&     1.06 (+0.04/-0.04)& 88.4&  0.97& 1998&------\\\hline

107& 8.81&26.13&  2&    33.06& 2797 (+275/-251)&     0.75 (+0.04/-0.03)& 88.3&  0.72& 2605&-1.51\\\hline

108& 8.71&24.79&  1&    42.75& 2435 (+149/-134)&     0.86 (+0.03/-0.04)& 89.4&  0.82& 2256&------\\\hline

109& 8.71&25.46&  1&    35.53& 2459 (+212/-196)&     0.68 (+0.03/-0.03)& 83.7&  0.80& 2281&------\\\hline

110& 9.12&27.02&  9&    36.01& 2866 (+275/-236)&     0.69 (+0.03/-0.02)& 82.4&  1.03& 2683&-1.59\\\hline

111& 8.04&24.64&  2&    48.25& 2480 (+134/-126)&     1.09 (+0.04/-0.03)& 87.9&  0.87& 2294&-1.17\\\hline

112& 8.04&25.46&  1&    37.72& 2413 (+181/-173)&     0.72 (+0.03/-0.03)& 86.1&  0.89& 2233&------\\\hline

113& 7.65&26.19&  3&    51.42& 2583 (+157/-141)&     0.74 (+0.02/-0.02)& 87.9&  0.79& 2412&------\\\hline

114& 7.26&24.74&  3&    36.79& 2370 (+134/-118)&     1.33 (+0.05/-0.05)& 86.4&  0.83& 2245&-0.94\\\hline

115& 7.37&25.46&  1&    40.42& 2521 (+188/-165)&     0.87 (+0.04/-0.03)& 87.5&  0.77& 2346&------\\\hline

116& 6.86&26.94&  4&    55.60& 2548 (+118/-110)&     0.89 (+0.02/-0.02)& 86.1&  0.86& 2375&------\\\hline

117& 6.55&25.46&  3&    39.74& 2585 (+157/-141)&     1.21 (+0.05/-0.04)& 88.5&  0.84& 2545&-0.83\\\hline

118& 6.49&26.13&  2&    49.78& 2388 (+126/-110)&     0.85 (+0.03/-0.03)& 88.7&  0.75& 2208&------\\\hline

119& 5.25&26.03&  5&    28.79& 2973 (+259/-228)&     1.49 (+0.08/-0.07)& 89.6&  0.86& 2837&------\\\hline

120& 5.79&26.80&  2&    43.83& 2462 (+157/-149)&     0.82 (+0.03/-0.03)& 89.1&  0.83& 2284&------\\\hline

121& 5.64&27.51&  3&    39.74& 2658 (+236/-212)&     0.79 (+0.03/-0.04)& 88.6&  0.83& 2492&------\\\hline

122& 4.32&27.29&  6&    41.85& 2788 (+173/-157)&     0.85 (+0.03/-0.03)& 88.7&  0.77& 2623&-1.18\\\hline

123& 3.69&28.42& 13&    39.42& 3144 (+259/-236)&     0.86 (+0.03/-0.04)& 89.7&  0.94& 3043&------\\\hline

124&23.96&12.97&  4&    53.26& 2242 (+102/ -94)&     0.75 (+0.01/-0.02)& 88.9&  1.05& 2062&------\\\hline

125&24.79&12.85&  2&    44.25& 2441 (+141/-134)&     0.62 (+0.02/-0.02)& 88.9&  0.91& 2262&------\\\hline

126&25.54&12.80&  5&    36.31& 2066 (+181/-157)&     0.44 (+0.01/-0.02)& 86.5&  1.04& 1891&------\\\hline

127&24.55&11.98&  5&    45.49& 2303 (+110/ -94)&     0.86 (+0.02/-0.02)& 86.2&  1.12& 2123&------\\\hline

128&25.46&11.76&  4&    43.53& 2268 (+118/-110)&     0.75 (+0.02/-0.02)& 88.3&  1.25& 2088&------\\\hline

129&26.17&11.74&  4&    47.47& 2356 (+126/-118)&     0.57 (+0.01/-0.02)& 88.8&  0.89& 2175&------\\\hline

130&27.12&11.17&  6&    43.58& 2191 (+134/-126)&     0.52 (+0.01/-0.02)& 88.4&  1.32& 2013&------\\\hline

131&26.58&10.62&  5&    42.18& 2303 (+118/-102)&     0.85 (+0.02/-0.02)& 84.6&  1.23& 2123&------\\\hline

132&28.25&10.08& 10&    34.23& 2472 (+157/-157)&     0.69 (+0.02/-0.02)& 85.4&  1.31& 2294&------\\\hline

133&27.47&10.05&  2&    25.62& 2191 (+157/-149)&     0.83 (+0.04/-0.03)& 88.3&  1.01& 2013&------\\\hline

\end{xtabular}
}

\end{document}